\documentclass[12pt,letterpaper]{article}
\usepackage{jheppub}
\usepackage{verbatim}
\usepackage{graphicx,times}
\usepackage{amsmath,amssymb,latexsym,float}
\usepackage{enumerate}

\numberwithin{equation}{section}

\newcommand{\abs}[1]{\left\vert#1\right\vert}

\newcommand{\R}{\text{\fontshape{n}\selectfont I\kern-.42exR}}

\newcommand{\1}{\text{\fontshape{n}\selectfont 1\kern-.56exl}}

\title{Resolving the scales of the Yang-Mills theory by means of an extra dimension}

\author[a,b]{Artan Bori\c{c}i}
\affiliation[a]{University of Tirana\\
             Blvd. King Zog I\\
             Tirana\\
             Albania
}
\affiliation[b]{Academy of Sciences of Albania\\
             Sq. Fan Noli\\
             Tirana\\
             Albania
}

\emailAdd{borici@fshn.edu.al}

\abstract{The Yang-Mills theory is part of the Standard Model of particle physics. The lack of the mathematical understanding of the theory stands out in theoretical physics. In order to address this problem we observe that a recently proposed general model beyond the Standard Model resolves the energy scales of a lattice regularized Yang-Mills theory by means of an extra dimension. The extra dimension ensures that all intermediate length scales of the physical system are available by definition. In this paper we study the role of the extra dimension also in the case of the free boson field. We find that if the extra dimension size is large, the model describes the classical motion of the system. In the opposite limit we recover its standard quantum mechanical motion without loss of information. Therefore, the Hilbert space of states in the presence of the extra dimension describes physical phenomena at all energy scales. This observation allows us to raise the description of length scales by an extra dimension at the level of a principle for the theories beyond the Standard Model, the only modeling constraint being the correspondence principle. As it was shown recently, the fermion ground state energy of a gauge invariant Hamilton operator of Dirac fermions gives a particular lattice regularization of the Yang-Mills theory. Integration of gauge fields gives a pure fermion theory of color singlet fermion-antifermion pairs at each lattice site evolving along the extra dimension. Color confinement follows directly from this property. It allows for a saddle point solution in the limit of a large number of colors. In this paper we find that the glueball spectrum of the Yang-Mills regularized theory is of the Hagedron type and bounded below by a positive value. We show also that the color charge is screened and the quark-antiquark potential is constant.
}

\begin{document}
\maketitle

\pagebreak

\section{Introduction}

The Yang-Mills theory is part of the Standard Model of particle physics. It is the basis of current understanding of the strong nuclear force. Its basic properties are asymptotic freedom and confinement of color. While asymptotic freedom is mathematically proven close to the continuum limit \cite{gross_wilczek,politzer}, linear confinement is rigorously shown to hold in the strong coupling limit of the Wilson regularized theory \cite{wilson,creutz}. Monte Carlo simulation of the latter has established the linear confinement as well as the computation of physical quantities with interest like masses of light hadrons and decay constants. While there are a lot of unanswered problems, lattice simulations remain the direct non-perturbative tool of investigation. As long as there is no mathematical understanding the only way forward are larger computers. The ADS/CFT correspondence \cite{maldacena} offers the possibility to understand gravity from field theory. Therefore, understanding the Yang-Mills theory is important for both strong force and gravity.

In order to address the solution of the theory we stay with the lattice regularization. A theory with a cutoff in place offers a mathematically precise definition of its continuum limit. In addition, we observe that the length scales of the theory may be resolved in the framework of a general model beyond the Standard Model by means of an extra dimension \cite{borici}. The extra dimension ensures that all intermediate length scales of the physical system are available by definition. In appendix \ref{free_boson_scales} we give another example, the one of the free boson field. If the extra dimension has a large size, the model describes the classical motion of the system. In the opposite limit one recovers the standard quantum mechanical motion of the same system. However, the Hilbert space of states is unitarily inequivalent to the Hilbert space obtained in the standard quantization. Therefore, the Hilbert space of states in the presence of the extra dimension describes physical phenomena at all energy scales.

These observations allow us to raise the extra dimension at the level of a principle dealing with the length scales of a physical system for the theories beyond the Standard Model. The only modeling constraint is the correspondence principle, i.e. the Standard Model should be derived by a theory which uses the extra dimension. In case of the Yang-Mills theory, the model adopted in reference \cite{borici} is a gauge invariant Hamilton operator of Dirac fermions in $d$ dimensions. There, it was shown that the fermion ground state energy of the model is a particular lattice regularization of the Yang-Mills theory. Moreover, it is was shown that the theory is written as a pure fermion theory of color singlet fermion-antifermion pairs at each lattice site evolving along the extra dimension. This property allows for a saddle point solution in the limit of a large number of colors. While there is no modeling recipe for the other sectors of the Standard Model, fermions and gauge invariance appear to by necessary ingredients.\footnote{We comment more on this matter at the end of the appendix \ref{free_boson_scales}.}

In the next section we give a short review the solution. In section \ref{eff_theory_properties} we discuss the properties of the saddle-point effective theory. Equations of motion are derived, the propagator is computed in detail and the spectrum of the saddle point theory is studied. The Yang-Mills theory glueball spectrum and the quark-antiquark potential are computed in section \ref{gluons}. In the last section we summarize and discuss the results. In appendix \ref{free_boson_scales} we discuss the extra dimension in the case of a free boson. In appendix \ref{action_to_ym} we compute the ground state energy of the theory using the action formulation.

\section{Solution review}
\label{solution_review}

In this section we set up notations and review the solution of the theory discussed in reference \cite{borici}. We begin the description with the Hamilton operator.

\subsection{Hamilton operator}

The Hamilton operator on a lattice with $d$ space-time dimensions and periodic boundary conditions is given by the staggered version of the lattice Dirac operator \cite{kogut_susskind} (more details are given in appendix \ref{action_to_ym})
\begin{equation}\label{Hamilton_operator}
\begin{aligned}
\hat H~~=&~~m\sum_{x,c}\hat\Psi(x)^*_c\hat{\gamma}_5(x)\hat\Psi(x)_c\\
+&~~\kappa\sum_{x,c,c'\mu}\hat{\gamma}_5(x)\eta_\mu(x)\left[\hat\Psi(x)^*_cU_{\mu}(x)_{cc'}\hat\Psi(x+\hat{\mu})_{c'}+\hat\Psi(x+\hat{\mu})^*_cU_{\mu}(x)^*_{cc'}\hat\Psi(x)_{c'}\right]\ ,
\end{aligned}
\end{equation}
where $\hat\Psi(x)^*_c,\hat\Psi(x)_c$ are fermion creation and annihilation operators of site $x$ and color component $c=1,2\ldots,N$, where $\hat\gamma_5(x)=\pm1$ depending on the parity of the lattice site $x$ and $\eta_1(x)=1,\eta_\mu(x)=(-1)^{x_1+\cdots+x_{\mu-1}},\mu=2,\ldots,d$. $m$ is the fermion mass and $U_{\mu}(x)_{cc'}$ are SU(N) matrix elements at each directed link $(x,x+\hat{\mu})$ on the lattice.\footnote{Matrices of the complex Gaussian ensemble will do as well. See appendix {\bf A} of \cite{borici}.} The coupling constant of the theory $\kappa>0$ is the strength of the hopping term. In the next subsection we deal with the emerging Yang-Mills theory from this formulation.

\subsection{Emergent Yang-Mills theory}

The Yang-Mills theory is derived from the Hilbert space trace
\begin{equation}
Z_F(U)=\text{tr}~e^{-N_\tau \hat H}\ ,
\end{equation}
where $N_\tau$ is the size of extra dimension. The effective action of the pure gauge theory is the ground state energy of the fermionic theory
\begin{equation}\label{ym1}
S_{\text{eff}}(U)=c_oN_\tau-c_1N_\tau\kappa^4\sum_{\mu\nu}\text{Tr~}U_\mu U_\nu U_\mu^*U_\nu^*+O(\kappa^6)+\text{h.c.}\ ,
\end{equation}
where the trace is taken in the tensor product space of the lattice sites and the SU(N) group, $c_o$ is real, $c_1=1/4$ and $U_\mu$ are the hopping matrices (see appendix \ref{action_to_ym} for the derivation)
\begin{equation}\label{hopping_matrices}
(U_\mu)_{xy;cc'}=U_\mu(x)_{cc'}\delta_{x+\hat\mu,y}\ .
\end{equation}
The first non-trivial term is the Wilson plaquette action if the length of the extra dimension is fixed by the relation
\begin{equation}\label{dimension_size_coupling_constant}
N_\tau=\frac{4}{\kappa^{4+\delta}}\ ,
\end{equation}
where $\delta$ is a positive integer. The rest of the terms are larger Wilson loops, which make the action long ranged. However, the contribution of the Wilson loop of length $2n$ decreases as a power of $\kappa^{2n-4-\delta}, n=2,3,\ldots$. Since the series converges for $\kappa\leq 1/(2d-1)$, the theory is local in the weak coupling limit. In this limit larger Wilson loops will not change the essence of the Yang-Mills theory. Asymptotic freedom and color confinement are properties of the theory.

The value of $\delta$ should be chosen such that only the Wilson term survives continuum limit. In section \ref{gluons} we show that Wilson loops are analytic functions of $\kappa^2$. In particular the mean value of the first term (see eq. (\ref{mean_plaquette}))
\begin{equation}\label{smallest_wilson_loop}
\sum_{\mu\nu}\langle\text{Tr~}U_\mu U_\nu U_\mu^*U_\nu^*\rangle=\text{\it w}_o-\text{\it w}_1\kappa^2+O(\kappa^4)\ ,
\end{equation}
where $\text{\it w}_o$ and $\text{\it w}_1$ are positive, gives a finite Wilson term if we set $\delta=2$. This way, the coupling constant $\kappa$ coincides, in the weak coupling limit, to the one of the Wilsonian theory and the length of extra dimension is fixed to the value
\begin{equation}
N_\tau=\frac{4}{\kappa^6}
\end{equation}
Note however that since the theory is solved by integrating first the gauge field the value of $\delta$ does not effect the structure of Green's functions of the theory. It does however effect the level spacing of the spectrum. Nonetheless, the value $\delta=1$ set in reference \cite{borici} is in error since with this value we may not relate the solution of the theory to that of the Yang-Mills theory.


The Dirac theory defined in equation (\ref{Hamilton_operator}) is a theory beyond the Standard Model where forces are absent. The emerging Yang-Mills theory is a general example that the fermion ground state induces a holonomy of gauge fields. The idea may be tested experimentally using the technique of trapped ions. If we put, for example, four fermions in a ring topology and gauge fields are $U(1)$ phase factors, we may define the Hamilton operator
\begin{equation}
H_\theta=c_1^*e^{i\theta_1}c_2+c_2^*e^{i\theta_2}c_3+c_3^*e^{i\theta_3}c_4+c_4^*e^{i\theta_4}c_1+h.c.\ ,
\end{equation}
where $c_j^*,c_j,j=1,2,3,4$ are fermion operators that satisfy anicommutation relations $\{c_j,c_l^*\}=\delta_{jl}$ and $\theta_j,j=1,2,3,4$ are the phases of the $U(1)$ field. In this case
we expect the emergence of a magnetic field perpendicular to the plane containing the ring. The exact experimental procedure is outside the scope of this paper. Next, we introduce the action of the theory.

\subsection{The action}

The action of the theory is\footnote{Gauge invariance along the extra dimension gives another model which deserves a separate study.}
\begin{equation}\label{I_action}
\begin{aligned}
{\cal I}~~~~=&~~~~\sum_{x,\tau,\tau',c}\bar{\psi}(x,\tau)_c\left[m\delta_{\tau,\tau'}+\hat{\gamma}_5(x)\hat{\partial}_t(\tau,\tau')\right]\psi(x,\tau')_c\\
&+\kappa\sum_{x,\tau,\mu,c,c'}\eta_\mu(x)\left[\bar{\psi}(x,\tau)_cU_{\mu}(x)_{cc'}\psi(x+\hat{\mu},\tau)_{c'}-\bar{\psi}(x+\hat{\mu},\tau)_cU_{\mu}(x)^*_{cc'}\psi(x,\tau)_c\right]\ ,
\end{aligned}
\end{equation}
where $\tau$ labels lattice sites along the extra dimension, $\psi(x,\tau)_c,\bar{\psi}(x,\tau)_c$ are fermion fields with color index $c$. They satisfy antiperiodic boundary conditions in $\tau$, whereas $\hat{\partial_\tau}$ is the lattice derivative, in our case, the symmetric differences matrix
\begin{equation}\label{lattice_derivative}
\hat{\partial_\tau}(\tau,\tau')=\frac{1}{2}\left(\delta_{\tau+1,\tau'}-\delta_{\tau-1,\tau'}\right)\ .
\end{equation}
Integration of gauge fields in the small $\kappa$ regime gives the pure fermion action in terms of color singlet fermion-antifermion pairs
\begin{equation}\label{pure_fermion}
\begin{aligned}
&S=\sum_{x,t,t',a}\bar{\psi}_a(x,t)\left[m\delta_{t,t'}+\hat{\gamma}_5(x)\hat{\partial}_t(t,t')\right]\psi(x,t')_a\\
&~~~~+N\sum_{x,\mu,t}F\left[-\frac{\kappa^2}{N^2}\sum_{t',a,b}\bar{\psi}(x,t)_b\psi(x,t')_b\bar{\psi}(x+\hat{\mu},t')_a\psi(x+\hat{\mu},t)_a\right]\ .
\end{aligned}
\end{equation}
Therefore, color confinement is trivial in this formulation. We use the first order expansion of $F(~.~)$ and have\footnote{As it is explained in \cite{borici}, the leading term dominates the solution.}
\begin{equation}\label{psi_action}
\begin{aligned}
&S=\sum_{x,\tau,\tau',c}\bar{\psi}(x,\tau)_c\left[m\delta_{\tau,\tau'}+\hat{\gamma}_5(x)\hat{\partial}_t(\tau,\tau')\right]\psi(x,\tau')_c\\
&~~~~+\frac{\kappa^2}{N}\sum_{x,\mu,\tau,\tau',c,c'}\bar{\psi}(x,\tau)_c\psi(x,\tau')_c\bar{\psi}(x+\hat{\mu},\tau')_{c'}\psi(x+\hat{\mu},\tau)_{c'}\ .
\end{aligned}
\end{equation}
Bosonization of fermions with the field $\Sigma(x,\tau,\tau')$ gives the action
\begin{equation}\label{sigma_action}
S_\Sigma=N\sum_{x,\tau}\left\{\ln\left[m+\hat{\gamma}_5(x)\hat\partial_\tau+\Sigma(x)\right]\right\}(\tau,\tau)-\frac{N}{2\kappa^2}\sum_{x,y,\tau,\tau'}\Sigma(x,\tau,\tau')(A^{-1})(x,y)\Sigma(y,\tau',\tau)\ ,
\end{equation}
where $A$ is the matrix $A(x,y)=\sum_{\mu}(\delta_{x+\hat\mu,y}+\delta_{x-\hat\mu,y})$. In the following we recall Green's function equalities of the theory.

\subsection{Green's functions}

The Green's functions of the theory are given by the expression
\begin{equation}\label{definition_greens_functions1}
\langle\psi(y,\tau')_a\bar\psi(x,\tau)_b\rangle_{\cal I}=G(x,\tau,\tau')\delta_{xy}\delta_{ab}\ ,
\end{equation}
where $G(x,\tau,\tau')$ is the solution of saddle point equations
\begin{equation}\label{saddle_point_equations}
G^{-1}(x,\tau',\tau)=m\delta_{\tau,\tau'}+\hat{\gamma}_5(x)\hat{\partial}_t(\tau,\tau')+\kappa^2\sum_{\mu}[G(x+\hat{\mu},\tau',\tau)+G(x-\hat{\mu},\tau',\tau)]\ .
\end{equation}
The $\Sigma$-field solution is
\begin{equation}\label{sigma_solution}
\Sigma(x,\tau',\tau)=\kappa^2\sum_{\mu}[G(x+\hat{\mu},\tau',\tau)+G(x-\hat{\mu},\tau',\tau)]\ .
\end{equation}
Note that the two-point fermion correlators are color singlets, i.e. they are invariant under color group transformations. They show that fermions do not propagate in space-time irrespective of their mass. This way, the net effect of the gauge field in a gauge invariant Dirac theory is to turn fermions static and colors into degenerate flavors of fermion-antifermion pairs. Two-point functions $G(x,\tau,\tau')$ are space-time local fields as well as correlators along the extra dimension. The four-point function of the theory is defined by the equation\footnote{In \cite{borici} it was derived for for $x\neq y$. This slight generalization uses two Wick contractions.}
\begin{equation}\label{four_point_function}
\langle\psi(x,\tau)_a\bar\psi(x,\tau')_a\psi(y,\tau')_b\bar\psi(y,\tau)_b\rangle_{\cal I}=G(x,\tau',\tau)G(y,\tau,\tau')(1-\delta_{xy}\delta_{ab})\ .
\end{equation}
There are other expressions for various combinations of space, time and color. Like the two-point function the four-point function is a color singlet quantity. It vanishes identically for equal spatial lattice sites and equal color. Therefore, the theory does not allow for two fermion-antifermion pairs of identical color at the same place. Next we recall the effective field solution of the theory.

\subsection{Effective field theory}

In order to solve the Green's function equalities one starts by Fourier transforming $G(x,\tau,\tau')$ in time and get the local field $\hat G(x,\omega)$ at each lattice site $x$ for each frequency $\omega$. Then, the solution is split in two pieces
\begin{equation}\label{ansatz}
\hat{G}(x,\omega)=e^{-i\theta(x,\omega)}\left[\tilde{G}_o(\omega)+\tilde{G}(x,\omega)\right]\ ,~~~~\theta(x,\omega)=\arg\left[m+\hat{\gamma}_5(x)i\sin\omega\right]
\end{equation}
where $\tilde{G}_o(\omega)$ is the site-independent piece of the solution, whereas $\tilde{G}(x,\omega)$ is a fluctuation. The substitution in the saddle point equations (\ref{saddle_point_equations}) gives
\begin{equation}\label{saddle_point_equations1}
\frac{1}{\tilde{G}_o(\omega)+\tilde{G}(x,\omega)}=\mu(\omega)+2d\kappa^2\tilde{G}_o(\omega)+\kappa^2\sum_{\mu}[\tilde{G}(x+\hat{\mu},\omega)+\tilde{G}(x-\hat{\mu},\omega)]
\end{equation}
with
\begin{equation}\label{mu_omega}
\mu(\omega)=\abs{m+i\sin\omega}\ .
\end{equation}
The field $\tilde{G}(x,\omega)$ is taken to be small compared to $\tilde{G}_o(\omega)$ and one finds \cite{borici}
\begin{equation}\label{tilde_G_o}
\frac{1}{\tilde{G}_o(\omega)}=\mu(\omega)+2d\kappa^2\tilde{G}_o(\omega)~~~~\Rightarrow~~~~\tilde{G}_o(\omega)=\frac{-\mu(\omega)+\sqrt{\mu(\omega)^2+8d\kappa^2}}{4d\kappa^2}
\end{equation}
as well as the linear equations
\begin{equation}\label{first_order_constraint}
\tilde{G}(x,\omega)+\kappa^2\tilde{G}_o(\omega)^2\sum_{\mu}[\tilde{G}(x+\hat{\mu},\omega)+\tilde{G}(x-\hat{\mu},\omega)]=0\ .
\end{equation}
These equations are valid as long as the neglected terms in the quadratic equations (\ref{saddle_point_equations1}) are small. The effective action of the field $\tilde{G}(x,\omega)$ is quadratic\cite{borici}
\begin{equation}\label{eff_action_G}
S_{\tilde{G}}=\frac{N\kappa^2}{2}\sum_{x,\omega}M(\omega)^2\tilde{G}(x,\omega)^2+\frac{N\kappa^2}{2}\sum_{x,\mu,\omega}\left[\tilde{G}(x+\hat{\mu},\omega)-\tilde{G}(x,\omega)\right]^2\ ,
\end{equation}
where
\begin{equation}\label{scalar_field_masses}
M(\omega)^2=\frac{1}{\kappa^2\tilde{G}_o(\omega)^2}-2d=\frac{\mu(\omega)^2+\mu(\omega)\sqrt{\mu(\omega)^2+8d\kappa^2}}{2\kappa^2}
\end{equation}
are the masses of the scalar fields. In the next section we study the properties of the effective field theory.

\section{Effective field properties}
\label{eff_theory_properties}

We start with the derivation of the equations of motion of the theory.

\subsection{Equations of motion}

The linear equations (\ref{first_order_constraint}) furnish the equation of motion for the field $\tilde{G}(x,\omega)$. Its Fourier space counterpart reads
\begin{equation}\label{equation_of_motion}
\left[1+2\kappa^2\tilde{G}_o(\omega)^2\sum_\mu\cos q_\mu\right]\tilde{G}_{\cal F}(q,\omega)=0\ ,
\end{equation}
where $\tilde{G}_{\cal F}(q,\omega)$ is the Fourier transformed field of $\tilde{G}(x,\omega)$
\begin{equation}
\tilde{G}(x,\omega)=\frac{1}{V}\sum_{p}{\tilde G}_{\cal F}(p,\omega)~e^{ipx}\ .
\end{equation}
This is a free scalar field on the lattice. Its momentum space expression is given by
\begin{equation}\label{momentum_solution}
\tilde{G}_{\cal F}(p,\omega)=\tilde{\tilde G}_{\cal F}(p,\omega)\delta_{f(p,\omega),0}\ ,
\end{equation}
where $\tilde{\tilde G}_{\cal F}(p,\omega)$ is a regular function of momenta and frequency and $f(p,\omega)=0$ is the dispersion relation of the theory. Its explicit form is
\begin{equation}\label{disp_rel1}
1+2\kappa^2\tilde{G}_o(\omega)^2\sum_\mu\cos q_\mu=0\ .
\end{equation}
However, the dispersion relation deriving from the effective action (\ref{eff_action_G}) is
\begin{equation}\label{disp_rel2}
1-2\kappa^2\tilde{G}_o(\omega)^2\sum_\mu\cos p_\mu=0\ .
\end{equation}
Therefore, we have two dispersion relations. Note that (\ref{disp_rel1}) may be derived from (\ref{disp_rel2}) if we take the origin of the momenta at the $(\pi,\pi,\ldots,\pi)$ corner of the Brillouin zone. Since there are two solutions for each site parity (see (\ref{ansatz})), we have a total of four solutions. The origin of this degeneracy is the staggered fermion formulation used in this model. Staggered fermions describe four degenerate flavors of fermions.\footnote{Had we used Wilson fermions we would have had one solution.} In the following we compute the propagator of the theory.

\subsection{Propagator}

We begin by writing the action (\ref{eff_action_G}) in terms of the new field
\begin{equation}\label{varphi}
\tilde{\varphi}(x,\omega)=\tilde{G}_o(\omega)^{-1}\tilde{G}(x,\omega)\ .
\end{equation}
It reads
\begin{equation}\label{phi_action}
S_{\tilde{\varphi}}=\frac{N}{2}\sum_{x,y,\omega}\left[\delta_{x,y}-\kappa^2\tilde{G}_o(\omega)^2\sum_\mu\left(\delta_{x+\hat\mu,y}+\delta_{x-\hat\mu,y}\right)\right]\tilde{\varphi}(x,\omega)\tilde{\varphi}(y,\omega)\ .
\end{equation}
The propagator of this field is
\begin{equation}\label{finite_volume_propagator}
\begin{aligned}
\left\langle\tilde{\varphi}(x,\omega)\tilde{\varphi}(0,\omega)\right\rangle_{S_{\tilde{\varphi}}}~~&=~~\frac{1}{N}\frac{1}{V}\sum_p\frac{e^{ipx}}{1-2\kappa^2\tilde{G}_o(\omega)^2\sum_\mu\cos p_\mu}\\
=\left[M(\omega)^2+2d\right]~&\frac{1}{N}\frac{1}{V}\sum_p\frac{e^{ipx}}{M(\omega)^2+2\sum_\mu(1-\cos p_\mu)}
\end{aligned}
\end{equation}
with $x_\mu=1,2,\ldots,N_\mu,\mu=1,2,\ldots,d$. In the infinite volume limit it is the integral
\begin{equation}\label{boson_correlator}
\tilde{C}(x,\omega)=\frac{1}{N}\left[M(\omega)^2+2d\right]\int\frac{d^dp}{(2\pi)^d}\frac{e^{ipx}}{M(\omega)^2+2\sum_\mu(1-\cos p_\mu)}\ .
\end{equation}
We use two approaches to compute the propagator as a function of the lattice sites.

\subsection*{Residue theorem}

The standard method to compute the propagator is the residue theorem. Introducing the momentum dependent mass
\begin{equation}
\nu(\vec{p},\omega)^2=M(\omega)^2+2\sum_{\mu=1}^{d-1}(1-\cos p_\mu)
\end{equation}
the propagator is written in the form
\begin{equation}
\tilde{C}(\vec{x},x_d,\omega)=\frac{1}{N}\left[M(\omega)^2+2d\right]\int\frac{d^{d-1}p}{(2\pi)^{d-1}}~e^{i\vec{p}~\vec{x}}~\int\frac{dp_d}{2\pi}\frac{e^{ip_dx_d}}{\nu(\vec{p},\omega)^2+2(1-\cos p_d)}\ .
\end{equation}
Using the Feynman contour for the integral over $p_d$ for $x_d>0$ we get
\begin{equation}
\tilde{C}(\vec{x},x_d,\omega)=\frac{1}{N}\left[M(\omega)^2+2d\right]\int\frac{d^{d-1}p}{(2\pi)^{d-1}}\frac{e^{i\vec{p}~\vec{x}-{\tilde\nu}(\vec{p},\omega)x_d}}{\sqrt{\nu(\vec{p},\omega)^2+4}}
\end{equation}
with
\begin{equation}
e^{\frac{1}{2}{\tilde\nu}(\vec{p},\omega)}=\frac{\nu(\vec{p},\omega)+\sqrt{\nu(\vec{p},\omega)^2+4}}{2}\ .
\end{equation}
For $\vec{x}=0,x_d=R$ and since $\nu(\vec{p},\omega)^2\geq M(\omega)^2$ we get the upper bound
\begin{equation}\label{upper_bound_propagator}
\tilde{C}(0,\ldots,0,R,\omega)\leq\frac{1}{N}\frac{M(\omega)^2+2d}{\sqrt{M(\omega)^2+4}}~e^{-{\tilde M}(\omega)R}
\end{equation}
with
\begin{equation}\label{mass_upper_bound_propagator}
e^{\frac{1}{2}{\tilde M}(\omega)}=\frac{M(\omega)+\sqrt{M(\omega)^2+4}}{2}\ .
\end{equation}
For small $M(\omega)$ and large $R$ the integral is dominated by small momenta. In this case its value approaches the upper bound with ${\tilde M}(\omega)\approx M(\omega)$ and $M(\omega)^2\approx 0$ and therefore we have
\begin{equation}\label{small_mass_propagator}
\tilde{C}(0,\ldots,0,R,\omega)\approx\frac{d}{N}~e^{-M(\omega)R}\ .
\end{equation}
In the following, we give another expression valid in the large mass limit, which is also the case of the vanishing $\kappa$ limit.

\subsection*{Modified Bessel functions of the first kind}

We write the denominator of (\ref{boson_correlator}) as an exponential integral
\begin{equation}\label{bessel_integral}
\tilde{C}(x,\omega)=\frac{1}{N}\left[M(\omega)^2+2d\right]\int_0^\infty du~e^{-M(\omega)^2u}\prod_{\mu=1}^d\int_{-\pi}^{\pi}\frac{dp_\mu}{2\pi}e^{ip_\mu x_\mu-2u(1-\cos p_\mu)}\ .
\end{equation}
The integral over momenta in the right hand side of (\ref{bessel_integral}) yields
\begin{equation}
\tilde{C}(x,\omega)=\frac{1}{N}\left[M(\omega)^2+2d\right]\int_0^\infty du~e^{-\left[M(\omega)^2+2d\right]u}\prod_{\mu=1}^d I_{x_\mu}\left[2u\right]\ ,
\end{equation}
where the modified Bessel functions of the first kind
\begin{equation}\label{modified_bessel_first_kind}
I_{x_\mu}(2u)=\frac{u^{x_\mu+2\cdot 0}}{0!(x_\mu+0)!}+\frac{u^{x_\mu+2\cdot 1}}{1!(x_\mu+1)!}+\frac{u^{x_\mu+2\cdot 2}}{2!(x_\mu+2)!}\cdots
\end{equation}
have been used. In the large mass $M(\omega)$ limit the leading term dominates and we find
\begin{equation}\label{prop_large_mass_bessel}
\tilde{C}(x,\omega)=\frac{1}{N}\frac{\left[\frac{1}{M(\omega)^2+2d}\right]^{x_1+x_2+\cdots+x_d}}{x_1!x_2!\ldots x_d!}(x_1+x_2+\cdots+x_d)!\ .
\end{equation}
As expected, the lattice propagator is not rotationally invariant. However, if we set $x_1=x_2=\cdots=x_d=R$ we can get an expression in the equal coordinates case. Using the approximation $x!\approx\sqrt{2\pi x}~x^x~e^{-x}$ we have 
\begin{equation}\label{prop_large_mass_equal_R}
\tilde{C}(R,\ldots,R,\omega)=\frac{\sqrt{d}}{N}~\frac{e^{-Rd~\ln\left[2+\frac{M(\omega)^2}{d}\right]}}{(2\pi R)^{\frac{d-1}{2}}}\ .
\end{equation}
In terms of $r=R\sqrt{d}$ we find
\begin{equation}
\tilde{C}\left(\frac{r}{\sqrt{d}},\ldots,\frac{r}{\sqrt{d}},\omega\right)=\frac{d^{\frac{d+1}{4}}}{N}~\frac{e^{-r\sqrt{d}~\ln\left[2+\frac{M(\omega)^2}{d}\right]}}{(2\pi r)^{\frac{d-1}{2}}}\ .
\end{equation}
Another useful case is when $x_1=x_2=\cdots=x_{d-1}=0$ and $x_d=R$ are substituted in (\ref{prop_large_mass_bessel})
\begin{equation}\label{prop_large_mass}
\tilde{C}(0,\ldots,0,R,\omega)=\frac{1}{N}\left[\frac{1}{M(\omega)^2+2d}\right]^R\ .
\end{equation}
Using (\ref{scalar_field_masses}) in the vanishing $\kappa$ limit we have $M(\omega)^2+2d\simeq\frac{\mu(\omega)^2}{\kappa^2}$ and therefore
\begin{equation}\label{prop_large_mass_small_kappa}
\tilde{C}(0,\ldots,0,R,\omega)\simeq\frac{1}{N}\left[\frac{\kappa^2}{\mu(\omega)^2}\right]^R\ .
\end{equation}
We get the same result from the expression (\ref{prop_large_mass_equal_R}) specialized in the vanishing coupling limit and $d=1$. In the next section we compute the mass spectrum of the saddle point theory.

\subsection{Hagedorn spectrum}
\label{hagedorn_spectrum}

Zero momentum energies are computed by averaging the propagator (\ref{boson_correlator}) over space coordinates
\begin{equation}
\tilde{\tilde{C}}(x_d,\omega)=\frac{1}{N_1N_2\ldots N_{d-1}}\sum_{x_1,x_2,\ldots,x_{d-1}}\tilde{C}(x,\omega)\ .
\end{equation}
In the infinite volume limit, the right hand side is the one-dimensional integral
\begin{equation}\label{1d_integral}
\tilde{\tilde{C}}(x_d,\omega)=\frac{1}{N}\int_{-\pi}^{\pi}\frac{dp_d}{2\pi}\frac{e^{ip_dx_d}}{1-2(d-1)\kappa^2\tilde{G}_o(\omega)^2-2\kappa^2\tilde{G}_o(\omega)^2\cos p_d}\ .
\end{equation}
Using the residue theorem and the Feynman contour as in the previous subsection one has
\begin{equation}\label{averaged_propagator}
\tilde{\tilde{C}}(x_d,\omega)=\frac{1}{N}\frac{M(\omega)^2+2d}{\sqrt{M(\omega)^2+4}}~e^{-{\tilde M}(\omega)x_d}\ .
\end{equation}
The spectrum is massive with spectral gap
\begin{equation}\label{spectral_gap}
{\tilde M}(0)=2\ln\left[\frac{M(0)+\sqrt{M(0)^2+4}}{2}\right]\ .
\end{equation}
For vanishing coupling the spectrum has the form (see (\ref{scalar_field_masses}))
\begin{equation}\label{spectral_gap}
{\tilde M}(\omega)=\ln\left(\frac{m^2+\sin^2\omega}{\kappa^2}\right)\ .
\end{equation}
Defining the mean exponential mass
\begin{equation}
e^{\overline{M}}=\frac{m^2+\overline{\sin^2\omega}}{\kappa^2}=\frac{m^2+\frac{1}{2}}{\kappa^2}
\end{equation}
and since $N_\tau=4/\kappa^6$ we find that the number of states grows exponentially with $\overline{M}$
\begin{equation}
N_\tau\sim e^{3\overline{M}}\ ,
\end{equation}
i.e. the mass spectrum is of the Hagedorn type. In the next section we discuss the properties of the Yang-Mills theory.

\section{Yang-Mills theory properties}
\label{gluons}

In this section we study the main properties of the Yang-Mills theory, the glueball spectrum and the quark-antiquark potential. Our calculations are based on the equivalent insertions in the path integral \cite{borici}
\begin{equation}\label{insertion}
U_{\mu}(x)_{ab}\longleftrightarrow\sum_\tau{\psi}(x,\tau)_a\bar{\psi}(x+\hat{\mu},\tau)_b
\end{equation}
as well as the use of Green's functions of the theory. For example, taking the expectation value of the right hand side and using the definition of Green's functions (\ref{definition_greens_functions1}) one has
\begin{equation}
\sum_\tau\left\langle{\psi}(x,\tau)_a\bar{\psi}(x+\hat{\mu},\tau)_b\right\rangle=\delta_{\hat{\mu},0}~\delta_{ab}~\sum_\tau G(x,\tau,\tau)\ .
\end{equation}
Since $\delta_{\hat{\mu},0}=0$ one finds $\left\langle U_{\mu}(x)_{ab}\right\rangle=0$, which is the Elitzur theorem. Before studying glueballs we go into more examples involving gauge-invariant operators, such as plaquette, Polyakov loop and the Polyakov loop correlator.

\subsection{Plaquette}

This subsection illustrates in some detail calculations involving the plaquette
\begin{equation}
\left\langle P_{\mu\nu}(x)\right\rangle=\frac{1}{N}\sum_{a,b,c,d}~\left\langle U_\mu(x)_{ab}U_\nu(x+\hat\mu)_{bc}U_\mu(x+\hat\nu)^*_{cd}U_\nu(x)^*_{da}\right\rangle\ .
\end{equation}
Using the fermion-antifermion insertion (\ref{insertion}) in the path integral we define the plaquette in terms of the fermion theory
\begin{equation}
\begin{aligned}
\left\langle P_{\mu\nu}(x)\right\rangle&=\frac{c_P}{N^4N_\tau}\sum_{\tau_1\tau_2\tau_3\tau_4,abcd}\left\langle\psi(x,\tau_1)_a\bar{\psi}(x+\hat{\mu},\tau_1)_b\psi(x+\hat{\mu},\tau_2)_b\bar{\psi}(x+\hat{\mu}+\hat{\nu},\tau_2)_c\right.\\
&\text{\hspace{3.5cm}}\left.\psi(x+\hat{\mu}+\hat{\nu},\tau_3)_c\bar{\psi}(x+\hat{\nu},\tau_3)_d\psi(x+\hat{\nu},\tau_4)_d\bar{\psi}(x,\tau_4)_a\right\rangle\ ,
\end{aligned}
\end{equation}
where $c_P$ is the plaquette normalization constant. Note that the normalization of Wilson loops computed in this way is arbitrary. We set $c_P$ to a value which gives $\left\langle P_{\mu\nu}(x)\right\rangle=1$ in the continuum limit. Using (\ref{definition_greens_functions1}) and taking the Fourier transform in the extra dimension one has
\begin{equation}
\left\langle P_{\mu\nu}(x)\right\rangle=\frac{c_P}{N_\tau}\sum_\omega\left\langle\hat{G}(x,\omega)\hat{G}(x+\hat\mu,\omega)\hat{G}(x+\hat\mu+\hat\nu,\omega)\hat{G}(x+\hat\nu,\omega)\right\rangle\ ,
\end{equation}
where the last expectation is taken with respect to the effective saddle point theory (\ref{phi_action}). Using the solution Ansatz (\ref{ansatz}) and the definition of the scalar field $\tilde\varphi$ (\ref{varphi}) one writes
\begin{equation}
\begin{aligned}
&\left\langle P_{\mu\nu}(x)\right\rangle\\
&=\frac{c_P}{N_\tau}\sum_\omega\tilde{G}_o(\omega)^4\left\langle\left[1+\tilde{\varphi}(x,\omega)\right]\left[1+\tilde{\varphi}(x+\hat\mu,\omega)\right]\left[1+\tilde{\varphi}(x+\hat\mu+\hat\nu,\omega)\right]\left[1+\tilde{\varphi}(x+\hat\nu,\omega)\right]\right\rangle\\
&=\frac{c_P}{N_\tau}\sum_\omega\tilde{G}_o(\omega)^4\left[1+4\tilde{C}(\hat\mu,\omega)+2\tilde{C}(\hat\mu+\hat\nu,\omega)\right]\ ,
\end{aligned}
\end{equation}
where the definition (\ref{boson_correlator}) of infinite volume boson propagators has been used. In the leading order of the large $N$ expansion and vanishing coupling constant propagators may be neglected and we get
\begin{equation}
\left\langle P_{\mu\nu}(x)\right\rangle=\frac{c_P}{N_\tau}\sum_\omega\left[\frac{1}{\mu(\omega)^4}-\frac{8d\kappa^2}{\mu(\omega)^6}+O(\kappa^4)\right]\ .
\end{equation}
Computing the sums by definite integrals
\begin{equation}
\frac{1}{N_\tau}\sum_\omega\frac{1}{\mu(\omega)^{2k}}\rightarrow\int_{-\pi}^{\pi}\frac{d\omega}{2\pi}\frac{1}{(1+\sin^2\omega)^k}\ ,~~~~~~~~k=2,3\ ,
\end{equation}
the expression takes the form
\begin{equation}
\left\langle P_{\mu\nu}(x)\right\rangle=c_P~\left[\frac{3}{4\sqrt{2}}-8d\kappa^2\frac{19}{32\sqrt{2}}+O(\kappa^4)\right]\ .
\end{equation}
Setting $c_P=\frac{4\sqrt{2}}{3}$ the mean plaquette is
\begin{equation}\label{mean_plaquette}
\left\langle P_{\mu\nu}(x)\right\rangle=1-8d\kappa^2\frac{19}{24}+O(\kappa^4)\ .
\end{equation}
In the next subsection we discuss the Polyakov loop.

\subsection{Polyakov loop}

The Polyakov loop is computed in reference \cite{borici}. Here we extend it in the next to leading order of the large $N$ expansion. In terms of Green's functions of the theory we define it to be
\begin{equation}
{\cal P}(\vec{x})=\sum_\omega\left\langle\prod_{x_d=1}^{N_d}\hat{G}(\vec{x},x_d,\omega)\right\rangle\ ,
\end{equation}
where $\vec{x}=(x_1,x_2,\ldots,x_{d-1})$ and $N_d$ is the length of the Euclidean time in lattice spacing units. Substituting the solution Ansatz (\ref{ansatz}) and using the definition of the scalar field $\tilde\varphi$ (\ref{varphi}) we have
\begin{equation}
\begin{aligned}
&{\cal P}(\vec{x})\\
&=\sum_\omega\tilde{G}_o(\omega)^{N_d}\left\langle\left[1+\tilde{\varphi}(\vec{x},1,\omega)\right]\left[1+\tilde{\varphi}(\vec{x},2,\omega)\right]\cdots\left[1+\tilde{\varphi}(\vec{x},N_d,\omega)\right]\right\rangle\\
&=\sum_\omega\tilde{G}_o(\omega)^{N_d}\left[1+(N_d-1)\tilde{C}(0,1,\omega)+\cdots+1\tilde{C}(0,N_d-1,\omega)+O\left(\tilde{C}^2\right)\right]\ .
\end{aligned}
\end{equation}
Since boson propagators fall off exponentially we stay with the next to leading term $(N_d-1)\tilde{C}(0,1,\omega)$ in the right hand side. Substituting $\tilde{C}(0,1,\omega)$ from (\ref{prop_large_mass_small_kappa}) and for large $N_d$ we get
\begin{equation}
{\cal P}(\vec{x})=\sum_\omega\tilde{G}_o(\omega)^{N_d}\left[1+\frac{N_d}{N}\frac{\kappa^2}{\mu(\omega)^2}\right]\ .
\end{equation}
In the large $N_d$ limit the sum is dominated by a few low frequencies. We approximate
\begin{equation}\label{polyakov_loop}
{\cal P}(\vec{x})\approx 2\tilde{G}_o(0)^{N_d}\left(1+\frac{N_d\kappa^2}{N}\right)\ ,
\end{equation}
where the factor two comes from the doubler. The free energy of the static quark is defined from the large $N_d$ exponential fall off of the Polyakov loop
\begin{equation}\label{af_o}
aF_o=-\ln \tilde{G}_o(0)-\frac{1}{N_d}\ln2-\frac{1}{N_d}\ln\left(1+\frac{N_d\kappa^2}{N}\right)\ ,
\end{equation}
where we have reintroduced the lattice spacing. Substituting $\tilde{G}_o(0)$ and sending $N_d$ to infinity we get
\begin{equation}
aF_o=2d\kappa^2+O(\kappa^4)\ .
\end{equation}
We take the continuum limit of the theory by keeping the free energy fixed as the coupling constant goes to zero. The renormalization group beta function \cite{borici}
\begin{equation}
\beta(\kappa)=-a\frac{d\kappa}{da}=\frac{a~(\partial F_o/\partial a)}{(\partial F_o/\partial\kappa)}=-\frac{\kappa}{2}+O(\kappa^3)
\end{equation}
shows that the theory is asymptotically free. In the following we study the glueball spectrum of the Yang-Mills theory.

\subsection{Glueball spectrum}

Glueball states are computed from the decay of the plaquette correlation functions. We study the correlations of the scalar glueball operator
\begin{equation}
S(x_d)=\sum_{\vec{x},kl} P_{kl}(\vec{x},x_d)\ ,
\end{equation}
where the sum is over all positions of the space-like plaquette. The connected correlator is defined by the expression
\begin{equation}
\begin{aligned}
\left\langle S(x_d)S(0)\right\rangle_c&=\sum_{\vec{x},\vec{y},kl,mn}\left\langle P_{kl}(\vec{x},x_d)P_{mn}(\vec{y},0)\right\rangle_c\\
&=~~\sum_{\vec{x},kl}\left\langle P_{kl}(\vec{x},x_d)P_{kl}(\vec{x},0)\right\rangle_c\ .
\end{aligned}
\end{equation}
In terms of Green's functions it is written in the form
\begin{equation}
\begin{aligned}
\left\langle S(x_d)S(0)\right\rangle_c&=\sum_{\omega,\vec{x},kl}\left\langle\hat{G}(\vec{x},x_d,\omega)\hat{G}(\vec{x}+{\hat k},x_d,\omega)\hat{G}(\vec{x}+{\hat k}+{\hat l},x_d,\omega)\hat{G}(\vec{x}+{\hat l},x_d,\omega)\right.\\
&\text{\hspace{2cm}}\left.\hat{G}(\vec{x},0,\omega)\hat{G}(\vec{x}+{\hat k},0,\omega)\hat{G}(\vec{x}+{\hat k}+{\hat l},0,\omega)\hat{G}(\vec{x}+{\hat l},0,\omega)\right\rangle_c\ .
\end{aligned}
\end{equation}
Using the solution Ansatz (\ref{ansatz}) and evaluating Wick's contractions of the scalar field $\tilde\varphi$ (\ref{varphi}) we find the leading term correlator
\begin{equation}
\left\langle S(x_d)S(0)\right\rangle_c\propto\sum_\omega\tilde{G}_o(\omega)^{4x_d}4\tilde{C}\left(\vec{0},x_d,\omega\right)\ .
\end{equation}
Using (\ref{prop_large_mass_small_kappa}) the glueball spectrum is given by studying the exponential decay of each term of the right hand side
\begin{equation}
aM_g(\omega)=8d\kappa^2+O(\kappa^4)+\frac{1}{x_d}\ln\frac{N}{4}+\ln\frac{\mu(\omega)^2}{\kappa^2}\ ,
\end{equation}
where we have reinstated the lattice spacing. In the large $x_d$ limit and vanishing coupling it is
\begin{equation}
aM_g(\omega)=\ln\left(\frac{m^2+\sin^2\omega}{\kappa^2}\right)\ .
\end{equation}
The lightest glueball mass
\begin{equation}\label{scalar_glueball_mass}
aM_g(0)=\ln\frac{m^2}{\kappa^2}
\end{equation}
is a twice degenerate state at $\omega=0$ and $\omega=\pi$. The mass gap of the Yang-Mills theory is infinitely large in the continuum limit. The same argument as in the end of subsection \ref{hagedorn_spectrum} shows that the mass spectrum of the theory is of the Hagedorn type. In the following we compute the quark-antiquark potential.

\subsection{Quark-antiquark potential}

We compute the energy of a quark-antiquark pair from the decay of Polyakov loop correlators located at $\vec{x}=(0,\ldots,0)$ and $\vec{y}=(0,\ldots,R)$
\begin{equation}\label{polyakov_loop_correlator}
C_{\cal P}(R,N_d)=\sum_\omega\left\langle\prod_{x_d=1}^{N_d}\hat{G}(0,\ldots,0,R,x_d,\omega)\prod_{y_d=1}^{N_d}\hat{G}(0,\ldots,0,0,y_d,\omega)\right\rangle_c
\end{equation}
where the subscript indicates the connected correlator. Using the solution Ansatz (\ref{ansatz}) and the definition of the scalar field $\tilde\varphi$ (\ref{varphi}) we have
\begin{equation}\label{connected_correlator}
\begin{aligned}
&C_{\cal P}(\vec{x})=\sum_\omega\tilde{G}_o(\omega)^{2N_d}\left\langle\left[1+\tilde{\varphi}(0,\ldots,0,R,1,\omega)\right]\cdots\left[1+\tilde{\varphi}(0,\ldots,0,R,N_d,\omega)\right]\right.\\
&\text{\hspace{4.1cm}}\left.\left[1+\tilde{\varphi}(0,\ldots,0,0,1,\omega)\right]\cdots\left[1+\tilde{\varphi}(0,\ldots,0,0,N_d,\omega)\right]\right\rangle_c\ .
\end{aligned}
\end{equation}
The leading term of the right hand side is
\begin{equation}
C_{\cal P}(R,N_d)\approx\sum_{\omega}\tilde{G}_o(\omega)^{2N_d}N_d\tilde{C}(0,\ldots,0,R,0,\omega)\ .
\end{equation}
For large $N_d$ the sum is dominated by the vanishing frequency term and its doubler at $\omega=\pi$. Using (\ref{prop_large_mass_small_kappa}) we have
\begin{equation}
C_{\cal P}(R,N_d)\propto\tilde{G}_o(0)^{2N_d}\frac{N_d}{N}\kappa^{2R}\ .
\end{equation}
Then, the quark-antiquark potential follows from the large $N_d$ limit of the exponential decay of the correlator
\begin{equation}\label{qq_potential_definition}
aV(R)=-\frac{1}{N_d}\ln C_{\cal P}(R,N_d)\propto 4d\kappa^2+O(\kappa^4)+\frac{1}{N_d}\ln\frac{N}{N_d}+\frac{R}{N_d}\ln\frac{1}{\kappa^2}\ .
\end{equation}
Sending $N_d$ to infinity the continuum limit potential is twice the free energy of the static quark
\begin{equation}
V(R)=2F_o\ .
\end{equation}
In the following we summarize and discuss the results obtanied in this paper.

\section{Summary and discussion}

We have shown that the Yang-Mills theory, as regularized in this paper, has a positive mass gap in the continuum limit. We have shown also that the number of glueball states grows exponentially with the glueball mass. The quarks are screened and the quark-antiquark potential is constant. These conclusions are the consequence of the solution of the theory in the limit of large number of colors, as reviewed in section \ref{solution_review}.

Color confinement follows from the color singlet property of pure fermion action of the theory. There is no linear confinement in the continuum limit. The string tension is zero in the infinite volume limit. Monte Carlo data with the Wilson regularization show that linear confinement is a property of the theory. In two dimensions, the exact solution of Gross and Witten with the Wilson action gives also a non-zero tring tension \cite{gross_witten}. The discrepancy may be explained by the presence of larger Wilson loops in the action: although vanishingly small in the continuum limit, they destroy the area law in the same way as dynamical quarks screen the linear potential of the Wilson action in the infinite volume QCD.

We studied also the continuum limit of the free energy of the static quark. The result does not change if one considers the quark-antiquark potential. These quantities give direct access to the coupling constant of the theory. We find that the renormalization group beta function vanishes linearly with $\kappa$, unlike the $\kappa^3$ behavior of the asymptotic perturbation theory of the Wilson theory. However, the exact solution with the Wilson action at $d=2$ shares the linear behavior of the beta function in the vanishing coupling limit. At $d=4$, there are no Monte Carlo data with the Wilson action at vanishingly small couplings.

The present regularization of the Yang-Mills theory in $d$ dimensions is the ground state of a Dirac theory beyond the Standard Model. The net effect of the gauge field is a color singlet fermion-antifermion theory in $d+1$ dimensions. These pairs, which are elementary fermions fixed at each lattice, form a color singlet composite field condensate. The fluctuations of the latter propagate in space-time and may be observed. The color confinement observed in Nature may be described in terms of such a fermion-antifermion condensate. The spectrum of the fluctuation field is the glueball spectrum of the Yang-Mills theory. The Hagerdorn type of the spectrum, which is otherwise obscure in the Wilson regularization, is an evidence that the model with an extra dimension resolves the multiple scales of the Yang-Mills theory.

\subsection*{Acknowledgement}

I thank my wife Mirela for frequent discussions related to this research and especially for the questions related to the physical meaning of the extra dimension. I thank Philippe de Forcrand for sending his comments on the first draft of the paper. Special thanks go to Michael Creutz for the correspondence related to various drafts of the paper and useful suggestions regarding multiple scales and the role of the extra dimension.

\appendix

\section{Multiple scales of a free boson}
\label{free_boson_scales}

The size of the extra dimension used in the regularization of the Yang-Mills theory is fixed by the coupling constant value. According to the renormalization group beta function, each value of the coupling constant corresponds to a given length scale. The evolution of the system along the extra dimension is the evolution to reach that scale. In doing so, the system steps along all intermediate scales and accesses all energy scales. Therefore, the extra dimension represents the length scales of the physical system. In this section we study another example, the free boson field. The field operator $\hat\phi(x)$ is a complex valued operator defined on lattice sites $x$ of a one dimensional lattice with periodic boundary conditions. Its Fourier space representation is
\begin{equation}
\hat\phi(x)=\frac{1}{\sqrt{N_1}}\sum_{k=1}^{N_1}a_k~e^{ip_kx}\ ,
\end{equation}
where ladder operators satisfy commutation relations
\begin{equation}
\left[a_k,a_{k'}^*\right]=\delta_{kk'}\ ,~~~~~~~~k,k'=1,2,\ldots,N_1\ ,
\end{equation}
$N_1$ is the number of lattice sites and $p_k$ are the lattice momenta $
p_k=\frac{2\pi k}{N_1}, k=1,2,\ldots,N_1$. The Hamilton operator reads
\begin{equation}
H=\frac{1}{2}\sum_{x=1}^{N_1}\left[2\hat\phi(x)^*\hat\phi(x)-\hat\phi(x)^*\hat\phi(x+1)-\hat\phi(x+1)^*\hat\phi(x)\right]+\left\{\hat\phi(x)\leftrightarrow\hat\phi(x)^*\right\}\ .
\end{equation}
In terms of ladder operators it is written in the form
\begin{equation}
H=E_o+2\sum_{k=1}^{N_1}\left(1-\cos p_k\right)a_k^*a_k\ .~~~~~~~~E_o=\sum_{k=1}^{N_1}\left(1-\cos p_k\right)\ .
\end{equation}
Operators $a_k$ annihilate the ground state with energy $E_o$. The system is coupled to the extra dimension of size $N_\tau$. The partition function of the theory is
\begin{equation}
Z(N_\tau)=\text{Tr~}_{\cal H}~e^{-N_\tau H}\ ,
\end{equation}
where the trace is taken in the Hilbert space of states $\cal H$. While in the interacting theories, the size of the extra dimension is related to the interaction strength, in a free theory it is an overall coupling. In the large $N_\tau$ limit the right hand side is dominated by the ground state, whereas in the small $N_\tau$ limit all energy states contribute. In the following we show the consequences of this property in the Green's functions of the theory.

We define a complex valued field $\phi(x,\tau)$ on a two-dimensional lattice with periodic boundary conditions. The action of the theory is
\begin{equation}
\begin{aligned}
S&=\frac{1}{2}\sum_{xt}\left[\phi(x,\tau)^*\phi(x,\tau+1)-\phi(x,\tau+1)^*\phi(x,\tau)\right]\\
&+\sum_{xt}\left[2\phi(x,\tau)^*\phi(x,\tau)-\phi(x,\tau)^*\phi(x+1,\tau)-\phi(x+1,\tau)^*\phi(x,\tau)\right]\ .
\end{aligned}
\end{equation}
We are interested in the infinite volume limit propagator
\begin{equation}
\langle\phi(x,\tau)\phi(0,0)^*\rangle=\int_{-\pi}^{\pi}\frac{dp}{2\pi}\int_{-\pi}^{\pi}\frac{d\omega}{2\pi}\frac{e^{i\omega\tau+ipx}}{i\sin\omega+2(1-\cos p)}\ .
\end{equation}
The first integral may be computed using the residue theorem for $\tau>0$. We find
\begin{equation}
\langle\phi(x,\tau)\phi(0,0)^*\rangle=\int_{-\pi}^{\pi}\frac{dp}{2\pi}\frac{e^{ipx-2\tau(1-\cos p)}}{\sqrt{1+4(1-\cos p)^2}}\ .
\end{equation}
As $\tau\rightarrow\infty$ the second integral allows the saddle point evaluation. In this case, low momenta dominate and the integral becomes Gaussian
\begin{equation}\label{correlator_xt}
\langle\phi(x,\tau)\phi(0,0)^*\rangle\approx\int_{-\infty}^{\infty}\frac{dp}{2\pi}e^{ipx-\tau p^2}=\frac{1}{\sqrt{4\pi\tau}}~e^{-\frac{x^2}{4\tau}}=\frac{1}{\sqrt{4\pi\tau}}-\frac{x^2}{\sqrt{16\pi\tau^3}}+O\left(\frac{x^4}{\sqrt{\tau^5}}\right)\ .
\end{equation}
This result shows that, in the large $\tau$ limit, the standard deviation of the field
\begin{equation}
\langle\phi(0,\tau)\phi(0,0)^*\rangle\approx\frac{1}{\sqrt{4\pi\tau}}
\end{equation}
is vanishingly small. In this limit, the expression on the right hand side of (\ref{correlator_xt}) shows also that the standard deviation of the derivative $\frac{d\phi}{dx}$ exists and vanishes in the large $\tau$ limit. Therefore, in the ground state, i.e. in the large $N_\tau$ limit, the pair of functions $\left\{\text{\footnotesize $\phi$},\frac{d\phi}{dx}\right\}$ may be used to define a classical theory, where $x$ has the meaning of physical time. This property is lost in the small $N_\tau$ limit. For example,  taking $N_\tau=1$ and since $\omega_l=\frac{2\pi l}{N_\tau}, l=1,2,\ldots,N_\tau$, we have $\omega=2\pi$. The propagator in this case is
\begin{equation}
\langle\phi(x,1)\phi(0,0)^*\rangle_{N_\tau=1}=\int_{-\pi}^{\pi}\frac{dp}{2\pi}\frac{e^{ipx}}{2(1-\cos p)}=1\ ,
\end{equation}
where the result is obtained using the residue theorem for $x>0$. Although the derivative $\frac{d\phi}{dx}$ is well defined and has vanishing standard deviation, the field standard deviation is finite.

\subsection*{Synthesis}

In the large $N_\tau$ limit the free boson behaves classically. In the opposite limit we get the expected behavior of the free field theory in one dimension, which describes a free quantum mechanical particle. In the formulation studied here, the motion of a free particle is classical if it is in the ground state. If it accesses all energy levels, which are otherwise obscure and not modeled when the extra dimension is missing, its motion is quantum mechanical. Therefore, classical and quantum mechanical behaviors of the same physical system are related by a unitary evolution along the analytically continued extra dimension. However, the Hilbert space of a free particle is spanned by plane waves, whereas the Hilbert space of the boson field is the product of harmonic oscillator Hilbert spaces. These are unitarily inequivalent spaces which share the same Green's function in the quantum regime of the latter space. Nonetheless, the free boson Hilbert space describes physical phenomena at all energy scales. At low energies the system behaves classically, whereas at high energies quantum mechanically. Therefore, for the system studied here, classical mechanics is a well defined limit of quantum mechanics without loss of information.

In principle, all other sectors of the Standard Model may be derived using the physical model in 3+1+1 dimensions. As a starting point one may replace the gauge field by a general matrix with Gaussian distribution entries, as it is done in appendix A of reference \cite{borici} for the derivation of the regularized Yang-Mills action. The axioms of quantum field theory, such as the Osterwalder and Schrader axioms \cite{osterwalder_schrader}, may, in principle, be extended without difficulties to the physical model in 3+1+1 dimensions.

\section{From action to Yang-Mills theory}
\label{action_to_ym}

In this section we expand definitions given in section \ref{solution_review} as well as give a new derivation of the Yang-Mills theory starting from the fermion action. We start with the Hamilton operator of the theory. Let $\hat\Psi_c(x),\hat\Psi_c(x)^*$, $c=1,2,\ldots,N$ be $N$ fermion annihilation and creation operators at each site $x=(x_1,x_2,\ldots,x_d)$ on a regular Euclidean lattice in $d$ dimensions acting on the Hilbert space ${\cal H}$. They obey the anticommutation relations
\begin{equation}
\left\{\hat\Psi_c(x)^*,\hat\Psi_{c'}(x')\right\}=\delta_{cc'}\delta_{xx'}\ .
\end{equation}
The lattice is finite and we assume it to be a torus with $V=N_1N_2\cdots N_d$ number of sites, where $N_1,N_2,\ldots N_d$ are the number of sites along each dimension. The Hamiltonian operator (\ref{Hamilton_operator}) used in this theory is the Kogut-Susskind Hamiltonian \cite{kogut_susskind}. In principle, any lattice formulation of Dirac fermions will do the job. We have chosen the staggered formulation because it is simple enough for our purpose. Here, $\hat{\gamma}_5$ is the lattice site parity operator taking $\pm1$ values on even/odd lattice sites, i.e. $\hat\gamma_5(x)=(-1)^{x_1+\cdots+x_d}$ and $\eta_1(x)=1,\eta_\mu(x)=(-1)^{x_1+\cdots+x_{\mu-1}},\mu=2,\ldots,d$. Both, $\hat\gamma_5$ and $\eta_\mu$ are diagonal matrices on the lattice. The Hamilton operator may be written in temrs of the Hermitian fermion matrix
\begin{equation}\label{fermion_matrix}
h=m\hat{\gamma}_5+\kappa \hat{\gamma}_5\sum_\mu\eta_\mu(U_\mu-U_\mu^*)\ ,
\end{equation}
where the hopping matrices $U_\mu$ (see \ref{hopping_matrices} for their definition) satisfy the following commutation relations
\begin{equation}
\hat\gamma_5\eta_\mu-\eta_\mu\hat\gamma_5=0\ ,~~~~~~~~\hat\gamma_5U_\mu+U_\mu\hat\gamma_5=0\ ,~~~~~~~~\eta_\mu U_\mu-U_\mu\eta_\mu=0\ .
\end{equation}
This way, we write
\begin{equation}
\hat H=\sum_{x,y,c,c'}\hat\Psi(x)_c^*h(x,y)_{cc'}\hat\Psi(y)_{c'}\ ,
\end{equation}
where $h(x,y)_{cc'}$ are the matrix elements of the $VN\times VN$ fermionic matrix $h$.

A new Hamilton operator may be derived by the action (\ref{I_action}). We do this in the following. Introducing Grassmann valued fermion fields $\Theta(x,\tau)=\psi(x,\tau)$ and $\bar{\Theta}(x,\tau)=\bar{\psi}(x,\tau)\hat{\gamma}_5(x)$ with antiperiodic boundary conditions along the extra direction the action is written in the form
\begin{equation}\label{action}
\tilde{{\cal I}}=\frac{1}{2}\sum_{x,\tau,c}\left[\bar{\Theta}(x,\tau)_c\Theta(x,\tau+1)_c-\bar{\Theta}(x,\tau+1)_c\Theta(x,\tau)_c\right]+\sum_{x,y,t,c,c'}\bar{\Theta}(x,\tau)_ch(x,y)_{cc'}\Theta(y,\tau)_{c'}\ .
\end{equation}
The partition function of the theory is the Berezin integral
\begin{equation}
\tilde{Z}_F(U)=\int\prod_{x,\tau,c}d\Theta(x,\tau)_cd\bar{\Theta}(x,\tau)_c~e^{\tilde{\cal I}}\ .
\end{equation}
One finds
\begin{equation}\label{5d_fermion}
\tilde{Z}_F(U)=\det
\begin{pmatrix} h & \frac{1}{2} &&&&\frac{1}{2}  \\
                -\frac{1}{2} &h & \frac{1}{2} &&&  \\
                & -\frac{1}{2} &h & \frac{1}{2} && \\
                && \ddots & \ddots & \ddots &  \\
                &&& -\frac{1}{2} &h & \frac{1}{2} \\
                -\frac{1}{2} &&&& -\frac{1}{2} &h
\end{pmatrix}\ ,
\end{equation}
where each entry is a block $VN\times VN$ matrix of the $VNN_\tau\times VNN_\tau$ matrix. Relabeling the last column as the first one and defining $2\times 2$ block matrices
\begin{equation}
A=
\begin{pmatrix} \frac{1}{2} & -h\\
                          0 & \frac{1}{2}
\end{pmatrix}\ ,~~~~~~~~
B=
\begin{pmatrix} \frac{1}{2} & 0 \\
                h           & \frac{1}{2}
\end{pmatrix}\ ,~~~~~~~~
C=
\begin{pmatrix} 1 & 0 \\
                0 & -1
\end{pmatrix}\ ,
\end{equation}
one has:
\begin{equation}\label{large_matrix}
\tilde{Z}_F(U)=(-1)^{VN}\det
\begin{pmatrix} AC &  B &        &        & \\
                   & -A & B      &        & \\
                   &    & \ddots & \ddots & \\
                   &    &        &  -A    & B \\
                BC &    &        &        & -A
\end{pmatrix}\ ,
\end{equation}
where we have assumed $N_\tau$ even. Finally, using the cyclic reduction of the resulting matrix determinant one obtains (modulo a constant factor)
\begin{equation}\label{reduced_determinant}
\tilde{Z}_F(U)=\det\left[1+\left(B^{-1}A\right)^{-\frac{N_\tau}{2}}\right]\ ,
\end{equation}
where
\begin{equation}
B^{-1}A=
\begin{pmatrix} 1   & -2h \\
                -2h & 1+4h^2
\end{pmatrix}\ .
\end{equation}
This matrix is brought to block diagonal form and we get
\begin{equation}\label{action_partition_function}
\tilde{Z}_F(U)=\det\left(1+e^{-N_\tau\tilde{h}}\right)\ ,
\end{equation}
with
\begin{equation}\label{tilde_h}
\tilde{h}=
\begin{bmatrix} \frac{1}{2}\ln\left(1+2h^2+2\sqrt{h^2+h^4}\right)   & 0 \\
                0 & -\frac{1}{2}\ln\left(1+2h^2+2\sqrt{h^2+h^4}\right)
\end{bmatrix}\ .
\end{equation}
Therefore, the corresponding Hamilton operator is defined by
\begin{equation}\label{hamilton_operator_from_action}
\tilde{H}=\sum_{x,y,a,b}\tilde\Psi(x)_a^*\tilde{h}(x,y)_{ab}\tilde\Psi(y)_b\ ,
\end{equation}
where $\tilde\Psi(x)_a$ is a doublet of two copies of $\hat\Psi(x)_a$ operators. The degeneracy comes form the symmetric differences in the approximation of lattice derivative, eq. (\ref{lattice_derivative}).

Finally, we show that both Hamilton operators defined in (\ref{Hamilton_operator}) and (\ref{hamilton_operator_from_action}) give Yang-Mills theories with the same Wilson loop expansion structure. The effective action of the theory may be written in terms of gauge fields only
\begin{equation}
\tilde{S}_{\text{eff}}(U)=-\frac{1}{2}\ln\tilde{Z}_F(U)\ ,
\end{equation}
where we have discounted by a factor of two the overcounting that comes from the degeneracy in the action formulation. The energy of the ground state of the Hamiltonian is computed as the large $N_\tau$ limit of the effective action. The largest fermion mass on a lattice is proportional to $m/a$ and we fix it to be exactly $1/a$. Then, form (\ref{fermion_matrix}) we have
\begin{equation}\label{h_o}
h^2=1+\kappa^2h_o^2\ ,~~~~~~~~~~~~h_o=\hat{\gamma}_5\sum_\mu\eta_\mu(U_\mu-U_\mu^*)\ .
\end{equation}
For large $N_\tau$ only one of the blocks of the $\tilde{h}$ matrix (\ref{tilde_h}) survives in the exponential function of (\ref{action_partition_function}). Therefore, using the identity $\det A=e^{\text{Tr}\log A}$ the effective action of the theory is
\begin{equation}\label{effective_action_h22}
\tilde{S}_{\text{eff}}(U)=-\frac{N_\tau}{4}~\text{Tr~}\ln\left(1+2h^2+2\sqrt{h^2+h^4}\right)\ ,
\end{equation}
where the trace is taken in the tensor product space of the lattice sites and the SU(N) group. Expanding the logarithm of the effective action (\ref{effective_action_h22}) and staying with the terms of the first order in $h$, we obtain the approximation
\begin{equation}\label{effective_action_h2}
S_{\text{eff}}(U)=-\frac{N_\tau}{2}~\text{Tr~}\sqrt{\1+\kappa^2h_o^2}\ .
\end{equation}
As shown in \cite{borici}, this expression may be computed directly from (\ref{Hamilton_operator}). Expanding the right hand side in powers of $\kappa^2$ we get
\begin{equation}\label{ym1}
S_{\text{eff}}(U)=c_oN_\tau-c_1N_\tau\kappa^4\sum_{\mu\nu}\text{Tr~}U_\mu U_\nu U_\mu^*U_\nu^*+O(\kappa^6)+\text{h.c.}\ ,
\end{equation}
where $c_o$ is real, $c_1=1/4=0.25$. Expanding directly (\ref{effective_action_h22}) in $\kappa^2$ one gets
\begin{equation}
\tilde{S}_{\text{eff}}(U)=\tilde{c}_oN_\tau-\tilde{c}_1N_\tau\kappa^4\sum_{\mu\nu}\text{Tr~}U_\mu U_\nu U_\mu^*U_\nu^*+O(\kappa^6)+\text{h.c.}
\end{equation}
with $\tilde{c}_1=3/(8\sqrt{2})=0.265\ldots$. While both formulations have the same Wilson loop structure, they differ in the coefficients multiplying the loops. The coefficient of the plaquette differs by $6\%$. Scaling the length of the extra dimension according to the relation
\begin{equation}\label{N_t_kappa}
c_1N_\tau=\frac{1}{\kappa^{4+\delta}}
\end{equation}
we get the effective theory
\begin{equation}\label{ym2}
S_{\text{eff}}(U)=c_oN_\tau-\frac{1}{\kappa^{\delta}}\sum_{\mu\nu}\text{Tr~}U_\mu U_\nu U_\mu^*U_\nu^*+O(\kappa^{2-\delta})+\text{h.c.}\ ,
\end{equation}
which is the Wilson theory enlarged by larger loops. Since the mean value of the first term may be written in the form (see eq. (\ref{mean_plaquette}))
\begin{equation}\label{smallest_wilson_loop}
\sum_{\mu\nu}\langle\text{Tr~}U_\mu U_\nu U_\mu^*U_\nu^*\rangle=\text{\it w}_o-\text{\it w}_1\kappa^2+O(\kappa^4)\ ,
\end{equation}
where $\text{\it w}_o$ and $\text{\it w}_1$ are positive, the action corresponds to the Yang-Mills action in the continuum limit if we set $\delta=2$. The other terms vanish in the continuum limit.

We derived the effective action for $N_\tau$ even. For odd $N_\tau$ one may follow a more general approach. Writing the action (\ref{action}) in terms of the Fourier modes
\begin{equation}\label{omega_k}
\omega_k=\frac{\pi}{N_\tau}(2k+1)\ ,~~~~~~~~k=1,2,\ldots,N_\tau\ ,
\end{equation}
which respect the boundary condition along the extra dimension and integrating Grassmann fields the partition function may be written as a product of determinants
\begin{equation}\label{general_expression}
\tilde{Z}_F(U)=\prod_{k=1}^{N_\tau}\det\left(h+i\sin\omega_k\right)\ .
\end{equation}
If $N_\tau$ is even and using (\ref{h_o}) the product takes the form
\begin{equation}
\tilde{Z}_F(U)=\prod_{k=1}^{N_\tau/2}\det\left(1+\sin^2\omega_k+\kappa^2h_o^2\right)
\end{equation}
with an extra factor $\det h$ in case $N_\tau$ is odd. We neglect this factor since it gives a small contribution (see the last subsection). Therefore the effective action is
\begin{equation}
\tilde{S}_{\text{eff}}(U)=-\frac{1}{2}\sum_{k=1}^{N_\tau/2}\text{Tr~}\ln\left(1+\frac{\kappa^2}{1+\sin^2\omega_k}h_o^2\right)\ .
\end{equation}
The Wilson loop expansion is derived by expanding the logarithm in $\kappa^2$
\begin{equation}
\tilde{S}_{\text{eff}}(U)\propto-\kappa^4\tilde{c}_1N_\tau\sum_{\mu\nu}\text{Tr~}U_\mu U_\nu U_\mu^*U_\nu^*+O(\kappa^6)+\text{h.c.}\ ,
\end{equation}
where $\tilde{c}_1$ is the definite integral
\begin{equation}
\tilde{c}_1=\frac{1}{N_\tau}\sum_{k=1}^{N_\tau/2}\frac{1}{(1+\sin^2\omega_k)^2}\approx\frac{1}{2}\int_{-\pi}^{\pi}\frac{d\omega}{2\pi}\frac{1}{(1+\sin^2\omega)^2}=\frac{3}{8\sqrt{2}}\ .
\end{equation}
In a similar fashion one may compute coefficients of larger loops. Both methods agree in the large $N_\tau$ limit. Note that the effective Yang-Mills theory described here is valid in the case of two or more dimensions. In the following we examine the one-dimensional case.

\subsection*{One-dimensional case}

This case is special since the action is given in terms of the Polyakov loop
\begin{equation}
S_{\text{eff}}^{d=1}(U)\sim \pm~N_\tau\kappa^{N_d}N_d~\text{Tr~}U_1(1)U_1(2)\cdots U_1(N_d)+\text{h.c.}\ ,
\end{equation}
where the sign is determined by the parity of $N_d$ and boundary conditions. Scaling the length of the extra dimension according to the relation
\begin{equation}
N_\tau\sim\frac{1}{\kappa^{N_d+2}}
\end{equation}
the action is of the form
\begin{equation}
S_{\text{eff}}^{d=1}(U)\sim \frac{N_d}{\kappa^2}~\text{Tr~}U_1(1)U_1(2)\cdots U_1(N_d)+\text{h.c.}\ .
\end{equation}
In the following we give a precise meaning of a weakly coupled regularization of the Yang-Mills theory.

\subsection*{A weakly coupled regularized theory}
\label{weakly_coupled_reg_theory}

In this subsection we show that the regularization of the Yang-Mills theory considered in this paper is valid at weak coupling only. In the case of strong coupling the theory differs considerably from the plaquette theory. If $N_\tau$ is a small integer, which corresponds to a large value of the coupling constant $\kappa\sim 1$, the expression of the effective action (\ref{effective_action_h2}) is no longer valid. In this case one has to use the general expression (\ref{general_expression}). For example, at $N_\tau=1$ one has
\begin{equation}
\tilde{Z}_F(U)=\det h
\end{equation}
and therefore
\begin{equation}
\tilde{S}_{\text{eff}}(U)=-\frac{1}{4}~\text{Tr~}\ln\left(\1+\kappa^2h_o^2\right)\ .
\end{equation}
Since $\kappa\sim 1$ the logarithm may not be expanded in convergent power series of $\kappa$. However, it may be expanded in terms of convergent series of another parameter if one notices that the spectral radius of $h_o$ is bounded by $2d$. Defining
\begin{equation}
h_1^2=h_o^2-2d^2\ ,~~~~~~~~\kappa_1^2=\frac{\kappa^2}{1+2d^2\kappa^2}
\end{equation}
we have
\begin{equation}
1+\kappa^2h_o^2=\frac{\kappa^2}{\kappa_1^2}(1+\kappa_1^2h_1^2)\ .
\end{equation}
The spectral radius of $\kappa_1^2h_1^2$ is bounded by one and we may expand
\begin{equation}
\tilde{S}_{\text{eff}}(U)\propto-\frac{1}{4}~\text{Tr~}\ln\left(\1+\kappa_1^2h_1^2\right)
\end{equation}
in a convergent series of $\kappa_1^2$. In this case the effective coupling of the plaquette is
\begin{equation}
\kappa^2_{\text{eff}}\sim\frac{1}{\kappa_1^4}\sim(2d^2)^2\ .
\end{equation}
Therefore, although the effective coupling of the plaquette is strong, we may not expand the effective theory in terms of $\kappa$. The situation is similar with larger Wilson loops. Repeating the exercise in the weak coupling limit
\begin{equation}
S_{\text{eff}}(U)=-\frac{N_\tau\kappa}{2\kappa_1}~\text{Tr~}\sqrt{\1+\kappa_1^2h_1^2}
\end{equation}
and expanding the right hand side in terms of $\kappa_1^2$ the effective coupling of the plaquette is
\begin{equation}
\kappa^2_{\text{eff}}\sim\frac{1}{\kappa_1^3N_\tau\kappa}\sim\left(\frac{1+2d^2\kappa^2}{\kappa^2}\right)^{\frac{3}{2}}\frac{1}{N_\tau\kappa}\ .
\end{equation}
For vanishing $\kappa$ we get
\begin{equation}
\kappa^2_{\text{eff}}\sim\frac{1}{N_\tau\kappa^4}\sim\kappa^2\ ,
\end{equation}
as expected. In this sense, the regularization presented in this paper is a weakly coupled regularized Yang-Mills theory.

\end{document}